\documentclass[prl,10pt,twocolumn,superscriptaddress,showpacs]{revtex4}
\usepackage{amsmath}
\usepackage{latexsym}
\usepackage{tabularx, booktabs}
\usepackage{amssymb}
\usepackage{graphics,epstopdf}
\usepackage{graphicx}
\usepackage[colorlinks=true, citecolor=blue, urlcolor=blue ]{hyperref}
\usepackage{float}
\usepackage{graphicx}
\usepackage{amsfonts}
\usepackage{color}
\begin{document}

\title{Quantum mechanical violation of macrorealism for large spin and its robustness against coarse-grained measurements}

\author{Shiladitya Mal}
\email{shiladitya.27@gmail.com}
\affiliation{S. N. Bose National Centre for Basic Sciences, Block JD, Sector III, Salt Lake, Kolkata 700 098, India}

\author{Debarshi Das}
\email{debarshidas@jcbose.ac.in}
\affiliation{Centre for Astroparticle Physics and Space Science (CAPSS), Bose Institute, Block EN, Sector V, Salt Lake, Kolkata 700 091, India}

\author{Dipankar Home}
\email{dhome@jcbose.ac.in}
\affiliation{Centre for Astroparticle Physics and Space Science (CAPSS), Bose Institute, Block EN, Sector V, Salt Lake, Kolkata 700 091, India}

\begin{abstract}
 For multilevel spin systems, robustness of the quantum mechanical (QM) violation of macrorealism (MR) with respect to coarse grained measurements is investigated using three different necessary conditions of MR, namely, the Leggett-Garg inequality (LGI), Wigner's form of the Leggett-Garg inequality (WLGI) and the condition of no-signalling in time (NSIT). It is shown that for dichotomic sharp measurements, in the asymptotic limit of spin, the algebraic maxima of the QM violations of all these three necessary conditions of MR are attained. Importantly, the QM violations of all these persist in that limit even for \textit{arbitrary} unsharp measurements, i.e. for any non zero value of the sharpness parameter characterizing the degree of fuzziness of the relevant measurements. We also find that when different measurement outcomes are clubbed into two groups for the sake of dichotomising the outcomes, the asymmetry/symmetry in the number of outcomes in the two groups, signifying the degree of coarse graining of measurements, has a crucial role in discerning quantum violation of MR. The results clearly demonstrate that classicality does not emerges in the asymptotic limit of spin, whatever be the unsharpness and degree of coarse graining of the measurements.
\end{abstract} 

\pacs{03.65.Ud}

\maketitle

\section{I. INTRODUCTION} 
One of the central concepts underpinning the classical world view in the macroscopic domain is the notion of macrorealism (MR) which is characterized by the following two assumptions: {\it Realism:} At any instant, irrespective of measurement, a system is in any one of the available definite states such that all its observable properties have definite values. {\it Noninvasive measurability (NIM):} It is possible, in principle, to determine which of the states the system is in, without affecting the state itself or the system's subsequent evolution. That the conjunction of these two assumptions is in conflict with QM was first shown by Leggett and Garg \cite{LGI1,LGI2}. This was achieved by deriving from these assumptions of a testable inequality involving time-separated correlation functions corresponding to successive measurement outcomes pertaining to a system whose state evolves in time. Such an inequality, known as the Leggett-Garg inequality (LGI), turns out to be incompatible with the relevant and testable QM predictions. Thus, LGI provides a necessary condition for MR, whose empirical violation would necessarily imply repudiation of MR. In recent years, investigations related to LGI have been acquiring considerable significance, as evidenced by a wide range of theoretical and experimental studies; see, for example, a recent comprehensive review \citep{emary}.\\

Against the above backdrop, it is noteworthy that, apart from LGI, of late, two more necessary conditions of MR have been proposed. One of them has been called Wigner's form of LGI (WLGI) \citep{saha}, and the other one is known as no-signalling in time (NSIT) \cite{beyond lgi}. WLGI can be regarded as a temporal version of Wigner's form of the local realist inequality \citep{Wigner,dipankar} that is derived as a testable algebraic consequence of the probabilistic form of MR. On the other hand, the NSIT condition is formulated as a statistical version of NIM to be satisfied by any macrorealist theory.\\

In this paper, for any value of spin pertaining to \textit{multilevel spin systems}, we focus on studying the robustness of the QM violations of the aforementioned necessary conditions of MR with respect to the measurement scheme introduced by Budroni and Emary \citep{bdroni} and considering fuzziness of the relevant measurements modelled in terms of what is known as  `unsharp' measurement \citep{busch, PB, busch2,busch3,busch4}. We have also adopted a more general approach to that measurement scheme by coarse graining different outcomes, well suitable for the purpose of studying quantum-classical transition in the macroscopic limit. Here we may note that, to date, there has been only a few studies on the QM violation of MR in the context of multilevel spin systems. One of these studies \citep{bdroni1} shows that for dichotomic measurements involving only two projectors that project a multilevel system onto one of the two possible subspaces, the maximum QM violation of LGI for \textit{any} dimensional system is the same as that for a qubit system. Another study \citep{bdroni} considers measurements involving projections onto individual levels and, interestingly, this study shows that the QM violation of LGI increases with an increasing value of the spin of the system under consideration, with the algebraic maximum of the violation being attained for infinitely large value of spin. In the other study \citep{mal}, how a suitable choice of the measurement scheme can lead to optimal violation of LGI for an arbitrary spin system has been shown, thus improving upon an earlier result \citep{KB}. Now, in the background of these studies, the key new results obtained in this paper are as follows:\\

$\bullet$ Similar to the case of LGI, the QM violations of WLGI and NSIT for multilevel spin systems pertaining to measurements that involve projections onto individual levels, too, increase with an increasing value of the spin and the algebraic maxima of their violations are attained when the spin becomes infinitely large.\\

$\bullet$ For modelling the effect of coarse graining of the measurements, when measurement outcomes of multilevel spin system are dichotomised by clubbing different outcomes into two groups, say `$+$' and `$-$', then asymmetry/symmetry in the number of outcomes in the two groups has a crucial role in discerning quantum violation of MR. However, if the measurements are considered to be sharp, then these two groups of outcomes can be sharply distinguished, which is in general not true in the macroscopic limit.\\

$\bullet$ We have, therefore, introduced unsharpness of the measurements to make the boundary between the two groups of outcomes imprecise. It is found that for arbitrarily large values of the spin, QM violations of all the three aforementioned necessary conditions of MR {\it{persist}} for \textit{arbitrary} unsharpness of measurements associated with lower degree of coarse graining of the relevant measurements. For extreme coarse grained measurements, while there exists a threshold unsharpness of the measurements above which no QM violation of LGI or WLGI occurs, the QM violation of NSIT persists for any amount of unsharpness of the measurements.\\

Significance of the above mentioned results stems from the following consideration. It may be recalled that among the various approaches suggested for addressing the issue of classical limit of QM \citep{book1, book2, book3, book4, book5, book6, book7, book8, book9, b10, b11, b12, b13}, there are two strands of prevalent wisdom that are relevant to the results obtained in this paper. One is that classical physics emerges from the predictions of QM in the so called `macroscopic' limit when either the system under consideration is of high dimensionality, for example, large spin system, or if a low dimensional system is of large mass, or if it involves large value of any other relevant parameter such as energy. The other is that classicality arises out of QM under the restriction of coarse grained measurements for which one can empirically resolve only those eigenvalues of a relevant observable that are sufficiently well separated; in other words, this view point stipulates that the limits of observability of quantum effects in an appropriate `macroscopic limit' determine the way the classicality emerges \citep{KB, KBB}.\\\\

As regards the first approach mentioned above, we note that counter-examples questioning it have been pointed out. For instance, in the case of the Bell-EPR scenario, it has been shown that quantum features in the sense of violating local realist inequalities, persist in the `macroscopic' limit such as for the large number of constituents of the entangled system~\citep{mer1}, or for the large dimensions of the constituents of the entangled system \citep{gp,gp1,M}. Further, for the Bell-EPR scenario, the QM violation of the relevant local realist inequalities seems to increase even in the limit of large numbers of particles and large magnitude of spins considered together~\citep{HM}. On the other hand, in the case of temporal correlations for which the violation of MR is probed through the violation of LGI, all the relevant studies mentioned earlier \citep{bdroni,mal,bdroni1} reveal that, irrespective of the nature of measurements, the QM violation of LGI persists for arbitrary large value of spin of the system under consideration.\\\\

As regards the second approach mentioned above, it has been shown \citep{KB, KBB} that for a
class of Hamiltonians governing the time evolution, if one goes into the limit of sufficiently large spins, but can experimentally only resolve eigenvalues which are separated by much more than the intrinsic quantum uncertainty, then the measurement outcomes appear to be consistent with that of classical laws. This is taken to suggest that classicality emerges out of QM under the restriction of coarse grained measurements. Along this line of research there had been a number of investigations giving more insight into the nature of coarse graining of the measurements and emergence of classicality or persistence of quantumness. In \citep{martini} micro-macro nonlocal correlation was established. Quantum violation of local realism has been shown \citep{jeong} for entangled thermal states with very low detection efficiency, i.e., for extreme coarse grained measurement available. Large amount of violation of Bell inequality has been obtained \citep{eye} with human eye as detector in a micro-macro experiment and this violation is robust against photon loss. Precise (non coarse grained) measurements are shown \citep{raeisi} to be essential for demonstrating quantum features at the mesoscopic or macroscopic level and observing nonlocality becoming more difficult with increasing system size. In \citep{jk}, it has been discussed that quantum-classical transition is forced to occur when measurement references are coarsened, while this is not the case when the final projection is coarsened. This particular result has been discussed in detail at end of Section V. There are also other recent works in this direction \citep{b28}.\\\\

Now, given the status of the above mentioned studies concerning the two approaches in question, we would like to stress that, to date, no study has investigated what happens if the question of emergence of classicality in the `macroscopic limit' is addressed for {\it {arbitrary}} values of the quantum number such as spin (including the asymptotic limit), \textit{in conjunction} with taking into account the effect of coarse graining of the relevant measurements. It is this void in the relevant literature that the present paper seeks to fill by examining the question of emergence of classicality in terms of the respective QM violations of all the three necessary conditions of MR, namely, LGI, WLGI and NSIT, modelling the effect of coarse graining of the relevant measurements and introducing the unsharpness of measurements characterized by a sharpness parameter. The striking result revealed by this study is that classicality does {\it{not}} emerge even in the asymptotic limit of spin by such coarse graining and unsharpness of measurements. Let us now broadly indicate the plan of the paper.\\\\

In the following Section, we explain the relevant features of the system under consideration (an arbitrary spin system in a uniform magnetic field), the specific type of measurement scheme used, its generalisation and the way the fuzziness of the measurements is modelled by unsharp measurement. In Section III, the key results obtained using LGI and WLGI are discussed, which is followed by Section IV  pertaining to NSIT. In Section V, the key results obtained using LGI, WLGI and NSIT by generalising the scheme by which different measurement outcomes are clubbed together into two different groups are discussed using projective measurement, as well as, using unsharp measurement. Finally, in the concluding Section VI, we elaborate a bit on the significance of the results obtained and also indicate the directions for future studies.

\section{II. SETTING UP OF THE MEASUREMENT CONTEXT}

Consider a QM spin $j$ system in a uniform magnetic field of magnitude $B_0$ along the $x$ direction. The relevant Hamiltonian is $(\hbar = 1)$:
\begin{equation}
\label{hh}
H = \Omega J_x
\end{equation}
where $\Omega$ is the angular precession frequency $(\propto B_0)$ and $J_x$ is the $x$ component of spin angular momentum.
Consider measurements of the $z$ component of spin ($J_z$) whose eigenvalues are denoted by $m$. The  measurement scheme used here \citep{bdroni} has the following features:\\\\
$\bullet$ The quantity $Q$ is such that $Q=-1$ when $m=-j$ and for any other value of $m$ ranging from $-j +1$ to $+j$, $Q = +1$. We will denote by $Q_i$ and $m_i$ the value of $Q$ and the outcome of $J_z$ measurement respectively at instant $t_i$. Thus $Q_i = +$ (i.e. $Q_i = +1$)  means $m_i = -j+1$ or, $-j+2$ or, ... $j-1$ or, $j$ and $Q_i = -$ (i.e. $Q_i = -1$) means $m_i = -j$. This grouping scheme of the measurement outcomes is used in Section III and IV.\\\\
$\bullet$ We initialize the system so that at $t$=$0$, the system is in the state $|-j;j\rangle$ where $|m;j\rangle$ denotes the eigen state of $J_z$ operator with eigenvalue $m$.\\\\
$\bullet$ Consider measurements of $Q$ at times $t_1$, $t_2$ and  $t_3$ $(t_1<t_2<t_3)$ \& set the measurement times  as $\Omega t_1 = \Pi $ and $\Omega(t_2 - t_1)$ = $\Omega(t_3 - t_2)$  = $\frac{\Pi}{2}$. For any $j$, this choice of measurement times may not give the maximum quantum violation of LGI, WLGI or NSIT. However, this choice suffices to give an idea about the nature of QM violations of the relevant inequalities for large $j$.\\\\
$\bullet$ We have also adopted a measurement scheme which is more general than described earlier and more natural in the context of emergence of classicality at the macroscopic limit with coarse grained measurement. This is described bellow:\\
$Q =-1$ for $m = -j, ..., -j+x $,\\
$Q=+1$ for $m = -j+x+1, ..., +j$, where $0 < x \leq $ integer part ($j$) and $x$ being integer.\\     The asymmetry in the number of measurement outcomes clubbed together decreases and, hence, the degree of coarse graining of the measurement increases with an increasing value of $x$. This generalised grouping scheme of the measurement outcomes is used in Section V. Here for $x =0$, the aforementioned scheme is reproduced. $x=$ integer part ($j$) denotes the most macroscopic grouping scheme in the sense of describing the perfect coarse graining of the measurements.
\\\\
 
Next, we use the notion of unsharp measurement in the context of treating fuzziness of the measurement.  Unsharp measurement, a form of positive operator valued measurement (POVM), is well studied in the quantum formalism. In ideal sharp measurement, the probability of obtaining a particular outcome, say $m$ in case of $J_z$ measurement, and the corresponding post-measurement state are determined by the projector $P_m = |m;j\rangle \langle m;j|$. On the other hand, in the case of unsharp measurement, the probability of an outcome and the corresponding post-measurement state are determined by the effect operator, which is defined as
\begin{equation}
\label{effect}
F_m = \lambda P_m + (1 - \lambda) \frac{\mathbb{I}}{d}
\end{equation}
where $\lambda$ is the sharpness parameter, where $0 \leq \lambda \leq 1$, $P_m$ is the projector onto the state $|m;j\rangle$,
$\mathbb{I}$ is the identity operator and $d$ is the dimension of the system (for spin $j$ system, $d = 2j+1$). Here $(1-\lambda)$ denotes the amount of white noise present in any unsharp measurement. 
Given the above specification of the effect operator, the probability of an outcome, say $m$, is given by $Tr(\rho F_m)$ for which the post-measurement state is given by, $(\sqrt{F_m} \rho \sqrt{F_m}^{\dagger}) / Tr(\rho F_m)$, $\rho$ being the state of the system on which measurement is done.

\section{III. ANALYSIS USING LGI AND WLGI}

For the purpose of this paper, we shall use the following form of 3-term LGI \citep{emary}:

\begin{equation}
\label{lgi}
K_{LGI}= C_{12} +C_{23} - C_{13} \leq 1
\end{equation}
where $C_{ij}$ = $\langle Q_i Q_j\rangle$ is the correlation function of the variable Q at two times $t_i$ and $t_j$.\\\\
As regards  WLGI, since this has been introduced only recently \citep{saha}, we briefly recapitulate its formulation before indicating the specific form of WLGI that we will be using in this paper. In the context of WLGI, the notion of realism implies the existence of overall joint probabilities $\rho (Q_1, Q_2, Q_3)$ pertaining to different combinations of definite values of observables or outcomes for the relevant measurements, while the assumption of NIM implies that the probabilities of such outcomes would be unaffected by measurements. Hence, by appropriate marginalization, the observable probabilities can be obtained. For example, the observable joint probability $P(Q_{1}-, Q_{2}+)$ of obtaining the outcomes -1 and +1 for the sequential measurements of Q at the instants $t_1$ and $t_2$, respectively, can be written as
\begin{equation}
P(Q_{1}-, Q_{2}+) = \sum_{Q_{3}=\pm 1} \rho(-, +, Q_3) \nonumber
\end{equation}
\begin{equation}
= \rho (-, +, +) + \rho(-, +, -)
\end{equation}
Writing similar expressions for the other measurable marginal joint probabilities $P(Q_{1}+, Q_{3}+)$ and $P(Q_{2}+, Q_{3}+)$, we get, for example, the following combination

\begin{equation}
P(Q_{1}+, Q_{3}+) + P(Q_{1}-, Q_{2}+) - P(Q_{2}+, Q_{3}+)\nonumber
\end{equation} 
\begin{equation}
\label{eq1}
= \rho (+, -, +) + \rho(-, +, -)
\end{equation}

Then, invoking non-negativity of the joint probabilities occuring on the RHS of Eq.(\ref{eq1}) the following form of WLGI is obtained in terms of three pairs of two-time joint probabilities.

\begin{equation}
\label{eqq}
K_{WLGI} = P(Q_{2}+, Q_{3}+) - P(Q_{1}-, Q_{2}+) - P(Q_{1}+, Q_{3}+)\leq 0
\end{equation}

Similarly, other forms of WLGI involving any number of pairs of two-time joint probabilities can be derived by using various combinations of the observable joint probabilities. Here we consider the specific form of the three term WLGI mentioned above (Eq.(\ref{eqq})).\\

\begin{center}
\begin{table}
\begin{tabular}{ |c|c|c| } 
 \hline
 \textit{\textbf{\textbf{j}}} & \textit{\textbf{($K_{LGI}-1$)}} & \textit{\textbf{($K_{WLGI}-0$)}} \\
\hline
\hline
 $1$ & $0.50$ & $0.44$ \\
 $10$ & $1.75$ & $0.87$ \\ 
 $100$ & $1.92$ & $0.96$ \\
 
 \hline
\end{tabular}
\caption{Table showing that the QM violations of LGI and WLGI increase with increasing values of the spin for ideal sharp measurement.} \label{tab1}
\end{table}
\end{center}

\emph{For projective measurement:} In order to calculate the expectation values and joint probabilities appearing in the aforementioned forms of LGI and WLGI, we proceed by writing the relevant time evolution operators as, for example, the time evolution operator from the initial time $t=0$ to the instant of first measurement $t=t_1$, $U(t_1 - 0) = e^{-i\pi J_x} = R^2$ (where $R$ = $e^{-i\frac{\pi}{2}J_x}$), and all the subsequent measurements are equispaced in time.
Typically, any joint probability, for example, $P(Q_{2}+, Q_{3}+)$ for a spin $j$ system is calculated  using the Wigner D-matrix formalism and is of the form given by,

\begin{equation}
P(Q_{2}+, Q_{3}+)
= 1 - \frac{(4j)!}{4^{2j} [(2j)!]^2} + \frac{1}{2^{4j}} - \frac{1}{2^{2j}}
\end{equation}

Using such expressions, both $K_{LGI}$ and $K_{WLGI}$ can be evaluated. We then obtain in Eq.(\ref{lgi}) and Eq.(\ref{eqq}) respectively
\begin{equation}
\label{klgi}
K_{LGI} = 3 + 4^{1 - 2 j} - 4^{1 - j} - \frac{2^{1 - 4 j} (4 j)!}{((2 j)!)^2}
\end{equation}

\begin{equation}
\label{kwlgi}
K_{WLGI} = 1 + 4^{-2 j} - 4^{-j} - \frac{4^{-2 j} (4 j)!}{((2 j)!)^2}
\end{equation}
QM violations of LGI and WLGI are quantified by $(K_{LGI} - 1)$ and $(K_{WLGI} - 0)$ respectively. It is found that both these violations increase with increasing values of $j$. Specific results showing this feature for $j=1, 10, 100$ are given in Table \ref{tab1}.

From Eqs. (\ref{klgi}) and (\ref{kwlgi}) it can be seen that for $j\rightarrow \infty$,
$(K_{LGI}-1) \rightarrow 2$ \citep{bdroni} and  $(K_{WLGI}-0) \rightarrow 1$. Thus, in both these cases, the algebraic maxima of both $K_{LGI}$ and $K_{WLGI}$ are attained for infinitely large spin value of the system under consideration.\\

\emph{For unsharp measurement:} Next, considering in the context of unsharp measurement, the expression of a typical joint probability distribution is of the form given by,

\begin{equation}
P(Q_1 +, Q_2 -) = \nonumber
\end{equation}
\begin{equation}
\label{ee}
 \sum_{k=-j+1}^{j} Tr[ F_{-j} U_{\Delta t_{2}} \sqrt{F_k} U_{\Delta t_{1}} \rho_i U_{\Delta t_{1}}^\dagger \sqrt{F_k}^\dagger U_{\Delta t_{2}}^\dagger] 
\end{equation}
where $\rho_i =$ initial state of the system $=|-j;j\rangle \langle -j;j|$, $U_{\Delta t_1} = U(t_1 - 0)$ and $U_{\Delta t_2} = U(t_2 - t_1)$.
\begin{center}
\begin{table}
\begin{tabular}{ |c|c|c| } 
 \hline
 \textit{\textbf{\textbf{}}} & \multicolumn{2}{|c|}{\textit{\textbf{Ranges of $\lambda$ for}}} \\
  \textit{\textbf{\textbf{j}}} & \multicolumn{2}{|c|}{\textit{\textbf{which the QM violation}}} \\
 \cline{2-3}
\textit{\textbf{\textbf{}}} & \textit{\textbf{of LGI persists
}} & \textit{\textbf{of WLGI persists}} \\
 \hline
 \hline
 $1$ & $(0.85,1]$ & $(0.71,1]$ \\
 $10$ & $(0.35,1]$ & $(0.28,1]$ \\ 
 $100$ & $(0.12,1]$ & $(0.08,1]$ \\
  
 \hline
\end{tabular}
\caption{Table showing that the ranges of $\lambda $ for which the QM violations of LGI and WLGI persist for different spin values increase with increasing values of spin.} \label{tab2}
\end{table}
\end{center}

Now, using the form of the effect operator defined earlier given by Eq.(\ref{effect}), Hamiltonian mentioned in Eq.(\ref{hh}) and using Wigner D Matrix formalism, one can obtain the joint probability pertaining to our measurement context as the following
\begin{equation}
P(Q_1 +, Q_2 -) = \nonumber
\end{equation}
\begin{equation}
\frac{x^2 \lambda}{2^{2j}} + 2x\lambda \sqrt{\frac{1 - \lambda}{2j + 1}} \frac{1}{2^{2j}} + \frac{\lambda(1 - \lambda)}{2j + 1} \frac{2j}{2^{2j}} + \frac{x^2 (1 - \lambda)}{2j + 1} \nonumber
\end{equation}
\begin{equation}
 + 2 x (\frac{1 - \lambda}{2j + 1})^{\frac{3}{2}} + (\frac{1 - \lambda}{2j + 1})^2 2j
\end{equation}
where $x = (\sqrt{\frac{2j\lambda + 1}{2j + 1}} - \sqrt{\frac{1 - \lambda}{2j + 1}})$. Using such joint probabilities, one can obtain

\begin{equation}
K_{LGI} = \frac{1}{(1 + 2 j)^2 ((2 j)!)^2} 16^{-j}((16^j + 2 (-2 + 16^j) \lambda^2 +  4 j^2  \nonumber
\end{equation}
\begin{equation}
(16^j - 4^{1 + j} \lambda + 2 (2 + 16^j) \lambda^2) - 4 \lambda (-2 + 4^j + 2 \sqrt{1 - \lambda} \nonumber
\end{equation}
\begin{equation}
\sqrt{1 + 2 j \lambda}  -  2^{1 + 2 j} \sqrt{1 - \lambda} \sqrt{1 + 2 j \lambda} + 16^j \sqrt{1 - \lambda} \sqrt{1 + 2 j \lambda}) - \nonumber
\end{equation}
\begin{equation} 
 4 j (16^j + 2 \lambda (-2 + 2^{1 + 2 j} - 2^{1 + 4 j} + 2 \sqrt{1 - \lambda} \sqrt{1 + 2 j \lambda} -  \nonumber
\end{equation}
\begin{equation}
2^{1 + 2 j} \sqrt{1 - \lambda} \sqrt{1 + 2 j \lambda} + 16^j \sqrt{1 - \lambda} \sqrt{1 + 2 j \lambda})))\nonumber
\end{equation}
\begin{equation}
((2 j)!)^2 + 2 (1 + 2 j) \lambda (-2 + \lambda - 2 j \lambda + 2 \sqrt{1 - \lambda} \sqrt{1 + 2 j \lambda}) (4 j)!)
\end{equation}

\begin{center}
\begin{table}
\begin{tabular}{ |c|c|c|c|c| } 
 \hline
 \textit{\textbf{\textbf{}}} &  \multicolumn{4}{|c|}{\textit{\textbf{The magnitude of QM violation}}} \\
 \cline{2-5}
  \textit{\textbf{\textbf{j}}} &  \multicolumn{2}{|c|}{\textit{\textbf{of LGI for}}}  &  \multicolumn{2}{|c|}{\textit{\textbf{of WLGI for}}} \\
 \cline{2-5}
 \textit{\textbf{\textbf{}}} & \textit{\textbf{$\lambda = 0.7$}} & \textit{\textbf{$\lambda = 0.5$}}  & \textit{\textbf{$\lambda = 0.7$}} & \textit{\textbf{$\lambda = 0.5$}} \\
\hline
\hline
 $10$ & $0.59$ & $0.19$ & $0.31$ & $0.12$\\
 $50$ & $0.80$ & $0.37$ & $0.40$ & $0.19$\\ 
 $100$ & $0.85$ & $0.41$ & $0.43$ & $0.21$\\ 
  
 \hline
\end{tabular}
\caption{Table showing the QM violations of LGI and WLGI for different spin values $j$ and different values of the sharpness parameter $\lambda$ of the measurement.} \label{tab3}
\end{table}
\end{center}

and
\begin{equation}
K_{WLGI} = \frac{1}{(1 + 2 j)^2 ((2 j)!)^2} 16^{-j} ((4 j^2 \lambda (-4^j + \lambda + 16^j \lambda) - \lambda   \nonumber
\end{equation}
\begin{equation}
(-2 + 4^j - 16^j + \lambda + 2 \sqrt{1 - \lambda} \sqrt{1 + 2 j \lambda} - 2^{1 + 2 j} \sqrt{1 - \lambda}  \nonumber
\end{equation}
\begin{equation}
\sqrt{1 + 2 j \lambda} + 2^{1 + 4 j} \sqrt{1 - \lambda} \sqrt{1 + 2 j \lambda}) - 2 j (16^j + \lambda (-2 + 2^{1 + 2 j} \nonumber
\end{equation}
\begin{equation}
- 3 (16^j) + 2 \sqrt{1 - \lambda} \sqrt{1 + 2 j \lambda} - 2^{1 + 2 j} \sqrt{1 - \lambda} \sqrt{1 + 2 j \lambda} \nonumber
\end{equation}
\begin{equation}
 + 2^{1 + 4 j} \sqrt{1 - \lambda} \sqrt{1 + 2 j \lambda}))) ((2 j)!)^2 + (1 + 2 j) \lambda (-2 + \lambda \nonumber
\end{equation}
\begin{equation}
 - 2 j \lambda + 2 \sqrt{1 - \lambda} \sqrt{1 + 2 j \lambda}) (4 j)!)
\end{equation}

Now, for a particular value of $j$, the ranges of $\lambda$ for which the QM violations of LGI and WLGI persist differ with the range for WLGI being greater than that for LGI. Moreover, the robustness of QM violations of both LGI and WLGI with respect to unsharpness of the measurement increase with increasing values of $j$. This is illustrated by the results given in Table \ref{tab2}, which indicate that the ranges of $\lambda$ for which the QM violations of LGI and WLGI persist increase with increasing values of $j$.\\

Most interestingly, for $j\rightarrow \infty$, we get, for the QM violations of LGI and WLGI
\begin{equation}
(K_{LGI}-1) \rightarrow 2\lambda^2
\end{equation}
and
\begin{equation}
(K_{WLGI}-0) \rightarrow \lambda^2
\end{equation}
which show that the ranges for which the QM violations of LGI and WLGI persist become equal to $(0,1]$. On the other hand, for any $j$, magnitude of the QM violation of LGI (WLGI) decreases for decreasing values of $\lambda$. This is illustrated by the results shown in Table \ref{tab3}.\\

Thus it is shown that if one adopts the type of measurement scheme used here, in the macrolimit characterized by infinitely large spin values as well as for any non-zero value of the sharpness parameter (i.e. for an arbitrary degree of fuzziness of the relevant measurement), the QM violation of MR persists for both LGI and WLGI.\\

\emph{For mixed initial state:} The above mentioned results are obtained for the aforementioned pure initial state. Let us investigate whether such kind of behaviour persists when initial state becomes mixed which is the more realistic situation involved in actually testing the macrolimit of quantum mechanics. Here, instead of taking pure initial state $|-j;j\rangle$ at $t$=$0$, we initialize the system so that at $t$=$0$, the system is in the state $\rho$ given by,
\begin{eqnarray}
\label{mixed}
 \rho = v |-j;j\rangle \langle -j;j| + (1-v) \frac{\mathbb{I}}{d}
\end{eqnarray}
\begin{center}
\begin{table}
\begin{tabular}{ |c|c|c| } 
 \hline
 \textit{\textbf{\textbf{$j$}}} & \textit{\textbf{$v_{th}$ for LGI
}} & \textit{\textbf{$v_{th}$ for WLGI}} \\
 \hline
 \hline
 $1$ & $0.571$ & $0.276$ \\
 $10$ & $0.098$ & $0.052$ \\ 
 $100$ & $0.010$ & $0.005$ \\
  
 \hline
\end{tabular}
\caption{Table showing that the threshold visibilities of LGI and WLGI decrease with increasing values of the spin for ideal sharp measurement.} \label{tab4}
\end{table}
\end{center}
where, $v$ is the visibility parameter which changes the pure state into a mixed state and $(1-v)$ denotes the amount of white noise present in the state $|-j;j\rangle$ ($0 \leq v \leq 1$), $d$ is the dimension of the system, $\frac{\mathbb{I}}{d}$ is the density matrix of completely mixed state of dimension $d$. The minimum values of $v$ for which QM violates different necessary conditions of MR signify the maximum amounts of white noise that can be present in the given state for the persistence of the QM violation of the relevant necessary condition of MR,  and this value of $v$ is known as the \textit{threshold visibility $(v_{th})$} pertaining to the given necessary condition of MR.\\

We take the Hamiltonian and choice of measurement times as described earlier and joint probabilities are evaluated for projective measurements. Using the Wigner D Matrix formalism, we obtain
\begin{eqnarray}
K_{LGI}= \frac{1}{((1 + 2 j) ((2 j)!)^2)}((16^{-j} ((2^{(3 + 2 j)} -  (16^{j})(3) + \nonumber
\end{eqnarray}
\begin{eqnarray}
(2)(2 - (3) 2^{(1 + 2 j)} + (3) (16^j)) v + 
      2 j (16^j + \nonumber
\end{eqnarray}
\begin{eqnarray}
2 (2 - 2^{(1 + 2 j)} + 16^j) v)) ((2 j)!)^2 - 
   2 (1 + 2 j) v (4 j)!)))
\end{eqnarray}
and
\begin{eqnarray}
K_{WLGI}= \frac{1}{((1 + 2 j) ((2 j)!)^2)}(16^{-j} ((-4^j (-2 + 4^j) +  \nonumber
\end{eqnarray}
\begin{eqnarray}
(1 + 2^{(1 + 4 j)} - (3) 4^j + 2 (1 - 4^j + 16^j) j) v) \nonumber
\end{eqnarray}
\begin{eqnarray}
((2 j)!)^2 - (1 + 2 j) v (4 j)!))
\end{eqnarray}
For a particular value of $j$, the threshold visibility of LGI and WLGI \textit{differ} with the threshold visibility for WLGI being \textit{smaller} than that for LGI, which signifies that for a particular $j$, QM violation of WLGI persists for a greater amount of mixedness compared to that of LGI. Moreover, the threshold visibilities of both LGI and WLGI \textit{decrease} with \textit{increasing} values of $j$. These results are shown in Table \ref{tab4}.\\ 

For any j, magnitudes of the QM violations of LGI or WLGI become \textit{smaller} for decreasing values of $v$, or increasing mixedness introduced in the initial state, while, for any fixed $v$ (fixed amount of mixedness introduced in the initial state), magnitudes of the QM violations of LGI or WLGI become \textit{larger} for increasing values of $j$. These results are shown in Table \ref{tab5}.
\begin{center}
\begin{table}
\begin{tabular}{ |c|c|c|c|c|c|c| } 
 \hline

 \textit{\textbf{\textbf{}}} & \multicolumn{6}{|c|}{\textit{\textbf{Magnitude of the QM violation of}}}\\
 \cline{2-7}
 \textit{\textbf{\textbf{j}}} & \multicolumn{3}{|c|}{\textit{\textbf{LGI for}}} & \multicolumn{3}{|c|}{\textit{\textbf{WLGI for}}} \\
 \cline{2-7}
 & $v = 0.8$ & $v = 0.6$ & $v = 0.4$ & $v = 0.8$ & $v = 0.6$ & $v = 0.4$ \\

\hline
\hline
 $1$ & $0.27$ & $0.03$ & No violation & $0.32$ & $0.20$ & $0.08$ \\
 $10$ & $1.36$ & $0.97$ & $0.58$ & $0.69$ & $0.51$ & $0.32$ \\
 $100$ & $1.53$ & $1.14$ & $0.76$ & $0.77$ & $0.57$ & $0.38$ \\
  
 \hline
\end{tabular}
\caption{Table showing the QM violations of LGI and WLGI for different spin values and different mixedness incorporated in the initial pure state with ideal sharp measurement.} \label{tab5}
\end{table}
\end{center}

We also find for $j \rightarrow \infty$, the QM violations of LGI and WLGI are given by
\begin{equation}
(K_{LGI}-1) \rightarrow 2 v
\end{equation}
and
\begin{equation}
(K_{WLGI}-0) \rightarrow v
\end{equation} 
These results clearly show that for very large $j$ and for any  amount of mixedness introduced in the initial state of the system, the QM violation of MR \textit{persists} using LGI or WLGI. 

\section{IV. ANALYSIS USING THE NSIT CONDITION}
According to the NSIT condition, the measurement outcome statistics for any observable at any instant is independent of whether any prior measurement has been performed. In order to study the above condition, let us consider a system whose time evolution occurs between two possible states. Probability of obtaining the outcome -1 for the measurement of a dichotomic observable Q at an instant, say, $t_3$ without any earlier measurement being performed is denoted by $P(Q_3 = -1)$.
NSIT requires that $P(Q_3 = -1)$ should remain unchanged even when an earlier measurement is made at $t_2$; i.e.,
\begin{center}
\begin{table}
\begin{tabular}{ |c|c| } 
 \hline
 \textit{\textbf{\textbf{j}}} & \textit{\textbf{Magnitude of the QM violation of NSIT}} \\
\hline
\hline
 $1$ & $0.63$ \\
 $10$ & $0.87$ \\ 
 $100$ & $0.96$ \\
  
 \hline
\end{tabular}
\caption{Table showing that the QM violation of NSIT increases with increasing values of the spin for ideal sharp measurement.} \label{tab6}
\end{table}
\end{center}

\begin{equation}
P(Q_3 = -1) - [P(Q_2 = +1, Q_3 = -1) \nonumber
\end{equation}
\begin{equation}
\label{eq3}
 + P(Q_2 = -1, Q_3 = -1)] = 0
\end{equation}

QM violation of NSIT is quantified by the non-vanishing value of the LHS of Eq.(\ref{eq3}).\\

\emph{For projective measurement:} For spin j system, for $H=\Omega J_x$, using the measurement scheme discussed in Section II and the choice of measurement times as well as of the initial condition mentioned there, we obtain, using the Wigner D Matrix formalism,\\
\begin{equation}
P(Q_3 = -1) - [P(Q_2 = +1, Q_3 = -1) \nonumber
\end{equation}
\begin{equation}
+ P(Q_2 = -1, Q_3 = -1)] \nonumber
\end{equation}
\begin{equation}
= 1 - \frac{(4j)!}{4^{2j} [(2j)!]^2}
\end{equation}

It is found that the QM violation of NSIT increases with increasing values of $j$. This is illustrated by the representative results given in Table \ref{tab6}.\\

For $j\rightarrow \infty$, the QM violation of NSIT $\rightarrow 1$, which is the algebraic maximum of the LHS of the NSIT condition.\\

\emph{For unsharp measurement:} For `unsharp measurement' defined in terms of the sharpness parameter $\lambda$ for spin j system described earlier, the LHS of Eq.(\ref{eq3}) becomes

\begin{equation}
P(Q_3 = -1) - P(Q_2 +, Q_3 -) - P(Q_2 -, Q_3 -) = \nonumber
\end{equation}
\begin{equation}
\frac{2^{-4 j} \lambda}{(1 + 
   2 j)((2 j)!)^2} [2 + (-1 + 2 j) \lambda - 
   2 \sqrt{1 - \lambda} \sqrt{1 + 2 j \lambda}] \nonumber
\end{equation}
\begin{equation}
 [16^j ((2 j)!)^2 - (4 j)!]
\end{equation}

\begin{center}
\begin{table}
\begin{tabular}{ |c|c|c|c| } 
 \hline
\textit{\textbf{\textbf{j}}} & \multicolumn{3}{|c|}{\textit{\textbf{Magnitude of the QM violation of NSIT}}}\\
 \cline{2-4}
\textit{\textbf{\textbf{}}} & \textit{\textbf{for $\lambda = 0.1$
}} & \textit{\textbf{for $\lambda = 0.5$}} & \textit{\textbf{for $\lambda = 0.8$}} \\
 \hline
 \hline
 $1$ & $0.0004$ & $0.0521$ & $0.2263$ \\
 $10$ & $0.0026$ & $0.1418$ & $0.4502$ \\ 
 $100$ & $0.0063$ & $0.2085$ & $0.5726$ \\
 $\rightarrow \infty$ & $0.0100$ & $0.2500$ & $0.6400$ \\
  
 \hline
\end{tabular}
\caption{Table showing the QM violations of NSIT for different spin values $j$ and different values of the sharpness parameter $\lambda$ of the measurement.} \label{tab7}
\end{table}
\end{center}

It is then found that for an arbitrary value of spin j, the QM violation of NSIT persists for any non-zero value of the sharpness parameter $\lambda$.\\\\
For $j\rightarrow \infty$, for any $\lambda$, LHS of the NSIT condition is given by\\
\begin{equation}
 P(Q_3 = -1) - [P(Q_2 = +1, Q_3 = -1)\nonumber
 \end{equation}
 \begin{equation}
+ P(Q_2 = -1, Q_3 = -1)] \rightarrow \lambda^2
\end{equation}

Thus, even in the `macrolimit' characterized by $j\rightarrow \infty$, for any non-zero value of $\lambda$, the QM violation of MR persists using NSIT.\\\\
For a given $\lambda$, the QM violations of NSIT for different spin $j$ systems differ, increasing with increasing values of $j$. Also, for a given $j$, magnitudes of the QM violations of NSIT increase with increasing values of $\lambda$, i.e., for increasing sharpness of the measurement. This is true even for infinitely large value of spin. These results are illustrated in Table \ref{tab7}.\\

\emph{For mixed initial state:} Now, instead of taking pure initial state $|-j;j\rangle$ at $t$=$0$, we initialize the system so that at $t$=$0$, the system is in the state $\rho$ given by Eq.(\ref{mixed}). We take the Hamiltonian and choice of measurement times as described earlier and joint probabilities are evaluated for projective measurements. Using the Wigner D Matrix formalism, we obtain,
\begin{eqnarray}
K_{NSIT}= v - \frac{2^{-4 j} v (4 j)!}{((2 j)!)^2}
\end{eqnarray}
For any j, magnitude of the QM violation of NSIT becomes \textit{smaller} for decreasing values of $v$, or increasing mixedness introduced in the initial state, while, for any fixed $v$ (fixed amount of mixedness introduced in the initial state), magnitude of the QM violation of NSIT becomes \textit{larger} for increasing values of $j$. These results are shown in Table \ref{tab8}.
\begin{center}
\begin{table}
\begin{tabular}{ |c|c|c|c| } 
 \hline
 \textit{\textbf{\textbf{j}}} & \multicolumn{3}{|c|}{\textit{\textbf{Magnitude of the QM violation of NSIT}}}\\
 \cline{2-4}
 & \textit{\textbf{\textbf{for $v = 0.8$}}} & \textit{\textbf{\textbf{for $v = 0.4$}}} & \textit{\textbf{\textbf{for $v = 0.2$}}} \\
\hline
\hline
 $1$ & $0.50$ & $0.25$ & $0.13$ \\
 $10$ & $0.70$ & $0.35$ & $0.17$ \\
 $100$ & $0.77$ & $0.38$ & $0.19$ \\
\hline
\end{tabular}
\caption{Table showing the QM violations of NSIT for different spin values and different mixedness incorporated in the initial pure state with ideal sharp measurement.} \label{tab8}
\end{table}
\end{center}

For any $j$ (including arbitrarily large value of $j$), threshold visibility of NSIT is $0$. So, interestingly, for  very large $j$ and for any  amount of mixedness introduced in the initial state of the system, the QM violation of MR \textit{persists} using LGI, WLGI or NSIT.\\

\section{V. Analysis using LGI, WLGI and NSIT for a more general grouping scheme of the measurement outcomes}

Here we generalise the scheme by which different measurement outcomes are clubbed together into two groups. In this case $Q =-1$ for $m = -j, ..., -j+x $, and $Q=+1$ for $m = -j+x+1, ..., +j$, where $0 < x \leq $ integer part($j$) and $x$ being integer. Here the degree of coarse graining of the measurement increases with increase in $x$. Any fixed value of $x$ denotes a particular grouping scheme.\\

We initialize the system so that at $t=0$, the system is in the state $|-j;j\rangle$. We take the aforementioned Hamiltonian and choices of measurement times. \\

\emph{For projective measurement:} Here joint probabilities appearing in the aforementioned particular form of LGI, WLGI or NSIT are calculated for ideal sharp measurement using Wigner D Matrix formalism.\\ 

From numerical results it is found that for any $j$ (also for arbitrarily large value), QM violation of LGI exists for $x \leq $ integer part($j-1$) and no violation occurs for $x = $ integer part($j$); whereas  QM violations of WLGI and NSIT exist for any value of $x$, where $x \leq $ integer part($j$). This indicates that QM violations of different necessary conditions of MR persist for very large degree of coarse graining of the measurement. However, the magnitudes of the violations become \textit{smaller} for increasing values of $x$, or increasing the degree of coarse graining of the measurement.\\

For a fixed and finite value of $x$, magnitudes of QM violations of LGI, WLGI or NSIT become \textit{larger} for increasing values of $j$. For arbitrarily large values of $j$ ($\frac{j}{x} >> 1$), magnitudes of QM violations of LGI, WLGI or NSIT approach their respective algebraic maxima. However, the QM violations of different necessary conditions of MR approach their respective algebraic maxima \textit{slowly} as one increases $x$. These results are shown in Table \ref{tab9}.\\

\begin{center}
\begin{table}
\begin{tabular}{|*{7}{c|}}
\hline
 & \multicolumn{6}{|c|}{\textit{\textbf{Magnitude of the QM Violation of}}}\\
\cline{2-7} 
\textit{\textbf{$j$}} & \multicolumn{2}{|c|}{\textit{\textbf{LGI for}}} & \multicolumn{2}{|c|}{\textit{\textbf{WLGI for}}} & \multicolumn{2}{|c|}{\textit{\textbf{NSIT for}}} \\ \cline{2-7}
 & \textit{\textbf{$x = 10$}} & \textit{\textbf{$x = 20$}} & \textit{\textbf{$x = 10$}} & \textit{\textbf{$x = 20$}} & \textit{\textbf{$x = 10$}} & \textit{\textbf{$x = 20$}} \\
 \hline 
 \hline
$40$ & $1.52$ & $1.32$ & $0.76$ & $0.66$ & $0.76$ & $0.66$\\
$60$ & $1.61$ & $1.46$ & $0.81$ & $0.73$ & $0.81$ & $0.72$\\
$80$ & $1.67$ & $1.53$ & $0.83$ & $0.77$ & $0.83$ & $0.76$\\
$100$ & $1.70$ & $1.58$ & $0.85$ & $0.79$ & $0.85$ & $0.79$\\
\hline
\end{tabular}
\caption{Table showing the QM Violations of LGI, WLGI and NSIT for different values of $j$ and $x$ with ideal sharp measurement.} \label{tab9}
\end{table}
\end{center}

\emph{For unsharp measurement:} Now, instead of projective measurement, let us employ unsharp measurement of spin-$z$ component observable. For this case the effect operators are defined as,
\begin{eqnarray}
F_m = \lambda P_m + (1 - \lambda) \frac{\mathbb{I}}{d}.
\end{eqnarray}
From numerical results it is observed that, for any $j$, the ranges of the sharpness parameter for which the QM violations of LGI and WLGI persist become \textit{smaller} for increasing values of $x$, or increasing the degree of coarse graining of the measurement. And for a fixed and finite value of $x$, the ranges of the sharpness parameter for which the QM violations of LGI and WLGI persist become \textit{larger} for increasing values of $j$ and for arbitrarily large values of $j$ ($\frac{j}{x} >> 1$), both the ranges approach $(0,1]$. However, these ranges approach $(0,1]$ \textit{slowly} as one increases $x$. This is shown in Table \ref{tab10}.\\

Interestingly, The range of the sharpness parameter for which the QM violation of NSIT persists for arbitrary values of $j$ (including arbitrarily large values of $j$) and arbitrary values of $x$ is $(0,1]$. This indicates that, surprisingly, for any particular scheme of branching of the outcomes (i.e. for a fixed value of $x$), for arbitrarily large values of $j$ ($\frac{j}{x} >> 1$), QM violations of all the necessary conditions of MR persist for almost any nonzero value of the sharpness parameter.\\

\begin{center}
\begin{table}
\begin{tabular}{|*{7}{c|}}
\hline
 & \multicolumn{6}{|c|}{\textit{\textbf{The range of sharpness parameter $(\lambda)$}}}\\
 & \multicolumn{6}{|c|}{\textit{\textbf{for which the QM violation of}}}\\
\cline{2-7} 
\textit{\textbf{$j$}} & \multicolumn{3}{|c|}{\textit{\textbf{LGI persists for}}} & \multicolumn{3}{|c|}{\textit{\textbf{WLGI persists for}}} \\ 
\cline{2-7}
 & \textit{\textbf{$x = 5$}} & \textit{\textbf{$x = 7$}} & \textit{\textbf{$x = 9$}} & \textit{\textbf{$x = 5$}} & \textit{\textbf{$x = 7$}} & \textit{\textbf{$x = 9$}} \\
 \hline 
 \hline
$10$ & $(0.64, 1]$ & $(0.75, 1]$ & $(0.92, 1]$ & $(0.53, 1]$ & $(0.61, 1]$ & $(0.72, 1]$\\
$20$ & $(0.49, 1]$ & $(0.55, 1]$ & $(0.59, 1]$ & $(0.40, 1]$ & $(0.44, 1]$ & $(0.48, 1]$\\
$30$ & $(0.42, 1]$ & $(0.47, 1]$ & $(0.51, 1]$ & $(0.33, 1]$ & $(0.37, 1]$ & $(0.40, 1]$\\
$40$ & $(0.38, 1]$ & $(0.42, 1]$ & $(0.45, 1]$ & $(0.29, 1]$ & $(0.33, 1]$ & $(0.36, 1]$\\
\hline
\end{tabular}
\caption{Table showing the ranges of the sharpness parameter for which QM violations of LGI, WLGI and NSIT persist for different values of $j$ and $x$.} \label{tab10}
\end{table}
\end{center}

{\em{Mixed state of different total spins:}} Here it should be mentioned that in a macroscopic sample consisting of many microscopic spins, total spin of the ensemble may not be precisely defined in general. In these cases, instead of having total spin number $j$, the resulting state would be a mixture of different total spin numbers varying from $0$ to $j$. For such a state using the measurement schemes discussed in Section II, the magnitudes of QM violations of all the aforementioned necessary conditions of MR will be less than that for the state having well defined total spin $j$. This is because for such a state, the magnitudes of the QM violations will be weighted average of the violations due to all possible total spin of the system varying from $0$ to $j$. Now, from Table \ref{tab1}, \ref{tab6} and \ref{tab9} it is clear that the magnitudes of the QM violations of LGI, WLGI and NSIT decrease with decreasing values of spin. Hence, the QM violation of any of these necessary conditions of MR for such a mixed state will have to be less than that for a system having precise total spin.\\\\

\emph{Coarsening of the measurement times}: Here it should be noted that in this work we have not considered the coarse graining of measurement times. For the sake of completeness, it needs to be mentioned that in \citep{jk} the effect of coarsening of the measurement times has been discussed in detail. It has been shown that when coarse graining at the level of measurement outcome fails to reproduce classical behaviour, coarsening of the measurement time, which is a particular example of measurement `reference' \citep{jk}, can reproduce it. In the context of temporal correlations, coarsening of measurement time means unsharpness of measurement time over a range whose effect is taken into account by considering measurement of a suitably averaged observable. For this kind of measurement, depending upon the Hamiltonian, there exists a threshold value of the unsharpness of measurement time $\Delta_{th}$ above which the QM violation of MR in terms of LGI disappears for any value of spin.

\section{VI. CONCLUDING DISCUSSION}
The grouping scheme of the measurement outcomes used in the first part of this paper was invoked by Budroni and Emary \citep{bdroni} to show that the magnitude of the QM violation of LGI increases with increasing values of the spin and approaches the algebraic maxima in the limit of arbitrarily large spin value. In this paper, contingent upon using this grouping scheme, it is shown that the aforementioned result holds for the WLGI and NSIT conditions of MR as well, and importantly, even in the context of unsharp measurement, the QM violation of MR persists in the arbitrarily large value of the spin for an arbitrary sharpness of the relevant measurement. These results hold good even when we generalise the grouping scheme. Here the clubbing of the measurement outcomes into two groups makes the measurement coarse grained. However, the boundary between the two groups of outcomes remains precise which is, in general, not true in the realization of the macrolimit. Employing, in conjunction, unsharp measurement makes this boundary also imprecise. Thus, simultaneously clubbing different measurement outcomes together and invoking unsharp measurement enables to describe in a more natural way the coarse graining of the measurements. It is, therefore, emphatically demonstrated that classicality  does not emerge for such coarse grained measurement even for arbitrarily large spin value of the system.\\\\
An interesting upshot of the results obtained in this paper is that, for any particular grouping scheme of the measurement outcomes, the range of the sharpness parameter for which the QM violation of WLGI persists is greater than that of LGI. This indicates that given a spin value, there is a range of the sharpness parameter for which the QM violation of MR can be tested using WLGI, but not in terms of LGI. Interestingly, the ranges of the sharpness parameter for which the QM violations of WLGI and LGI persist increase with increasing values of the spin. On the other hand, the QM violation of NSIT persists for any non-zero value of the sharpness parameter for any arbitrary spin value.\\\\
Finally, we recall that in the investigations of this paper we have used the model of the unsharp measurement that defines the effect operator in terms of a single parameter, i.e. the `sharpness parameter' \citep{busch, PB, busch2, busch3, busch4}. A possible alternative way to model the unsharp measurement is to define the effect operator in terms of what is called as `biasness parameter' \citep{bias, bias2} along with the `sharpness parameter'. It should be instructive to investigate to what extent the results obtained in this paper would be affected by using such two-parameter model of the unsharp measurement. Another line of future investigation would be to explore the implications of coarse graining of the measurement times mentioned in \citep{jk} in the context of the work presented in this paper.

\section{VII. ACKNOWLEDGEMENTS}
The authors acknowledge the anonymous referee for valuable comments. DD acknowledges the financial support from University Grants Commission (UGC), Govt. of India. The research of DH is supported by the Dept. of Science and Technology (DST), Govt. of India and Center for Science, Kolkata.

\end{document}